\documentclass[11pt,twoside]{article}
\usepackage{asp2010}

\resetcounters

\begin{document}

\title{Seismic analysis of four solar-like stars observed during more than eight months by {\it Kepler}}
\author{S. Mathur$^{*,1}$, T.L. Campante$^{2,3}$, R. Handberg$^{3}$,  R.A. Garc\'ia$^{4}$, T. Appourchaux$^{5}$, T.R. Bedding$^{6}$, B. Mosser$^{7}$, W.J. Chaplin$^{8}$, J. Ballot$^{9,10}$, O. Benomar$^{5}$, A.~Bonanno$^{11}$, E. Corsaro$^{11}$, P. Gaulme$^{5}$, S. Hekker$^{8,12}$,  C. R\'egulo$^{13,14}$, D.~Salabert$^{15}$, G. Verner$^{8}$,  T.R. White$^{6,16}$,  I.M. Brand\~ao$^{17}$, O.L. Creevey$^{15}$, G.~Do\u{g}an$^{3}$,  M. Bazot$^{2}$, M.~S. Cunha$^{2}$, Y. Elsworth$^{8}$, D. Huber$^{6}$,  S.J. Hale$^{8}$, G. Houdek$^{18}$, C. Karoff$^{3}$, M. Lundkvist$^{3}$,T.S.~Metcalfe$^{1}$, J. Molenda-\.Zakowicz$^{19}$, M.J.P.F.G. Monteiro$^{3}$, M.J. Thompson$^{1}$, D. Stello$^{6}$, J.~Christensen-Dalsgaard$^{1,3}$, R.L. Gilliland$^{20}$, S.~D. Kawaler$^{21}$, H.~Kjeldsen$^{3}$, B.~D. Clarke$^{22}$, F.~R. Girouard$^{23}$, J.~R. Hall$^{23}$, E.V. Quintana$^{22}$, D.T. Sanderfer$^{23}$, and S.E. Seader$^{22}$\\
{$^*$} Affiliations are given at the end of the paper}

\begin{abstract}
Having started science operations in May 2009, the {\it Kepler} photometer has been able to provide exquisite data of solar-like stars. Five out of the 42 stars observed continuously  during the survey phase show evidence of oscillations, even though they are rather faint (magnitudes from 10.5 to 12). In this paper, we present an overview of the results of the seismic analysis of 4 of these stars observed during more than eight months. 
\end{abstract}

\section{Introduction}
Asteroseismology is the only method that allows us to pierce into stars and thus to obtain information on the structure and dynamics of their interiors
\citep{2010ApJ...723.1583M}. With the advent of space-based missions performing asteroseismic investigations, during the last decade we have witnessed the rise of a new era in the field of Asteroseismology. The most powerful satellite  in operation so far is the NASA {\it Kepler mission} \citep{2010Sci...327..977B}, an exoplanet hunting mission whose long stability and high-precision photometry can also be used to perform asteroseismic investigations \citep{2010PASP..122..131G}, in particular, for the study of solar-like stars \citep{2011Sci...332..213C}. {\it Kepler} will observe $\sim$~150,000 stars in the same field during at least 3.5 years, allowing us to also study stellar cyclic variations \citep{2010Sci...329.1032G}. Asteroseismic investigations are organized around the {\it Kepler} Asteroseismic Science Consortium \citep[KASC,][]{2010AN....331..966K} that decided to perform a survey phase of the solar-like stars, during the first year of operations, by changing the set of observed stars every month. However, 42 stars were continuously monitored in order to test and validate the time series photometry. Some of them showed solar-like oscillations. In this work, we give an overview of the results we have obtained on four of them, two F-type stars (KIC~11234888 and KIC~10273246) and two G-type stars (KIC~11395018, and KIC~10920273). We refer to \citet{2011ApJ...733...95M} and \citet{Campante2011} for a detailed analysis of these stars.

\section{Data Processing \& Results}

Short cadence (58.85s) time series \citep{2010ApJ...713L.160G} of the four stars were obtained by the {\it Kepler} photometer from May 2009 to March 2010. Therefore, $\sim$~320 days of continuous observations with a duty cycle higher than 90$\%$ were available for seismic investigations, which is a premiere in this research field. Due to the loss of all the outputs of the third CCD-module on January 9, 2010, only $\sim$~250 days were available for two of the four stars: KIC~11234888 and KIC~11395018. The raw light curves were corrected for three types of instrumental perturbations: outliers, jumps, and drifts, following the methods described in \citet{2011MNRAS.414L...6G}. The data of each {\it Kepler} quarter were concatenated after equalizing their mean values by fitting a 6th order polynomial to each segment. To remove the low-frequency instrumental trends, a high-pass filter was applied. 

\begin{figure*}[htbp]
\begin{center}
\includegraphics[width=4cm, angle=90]{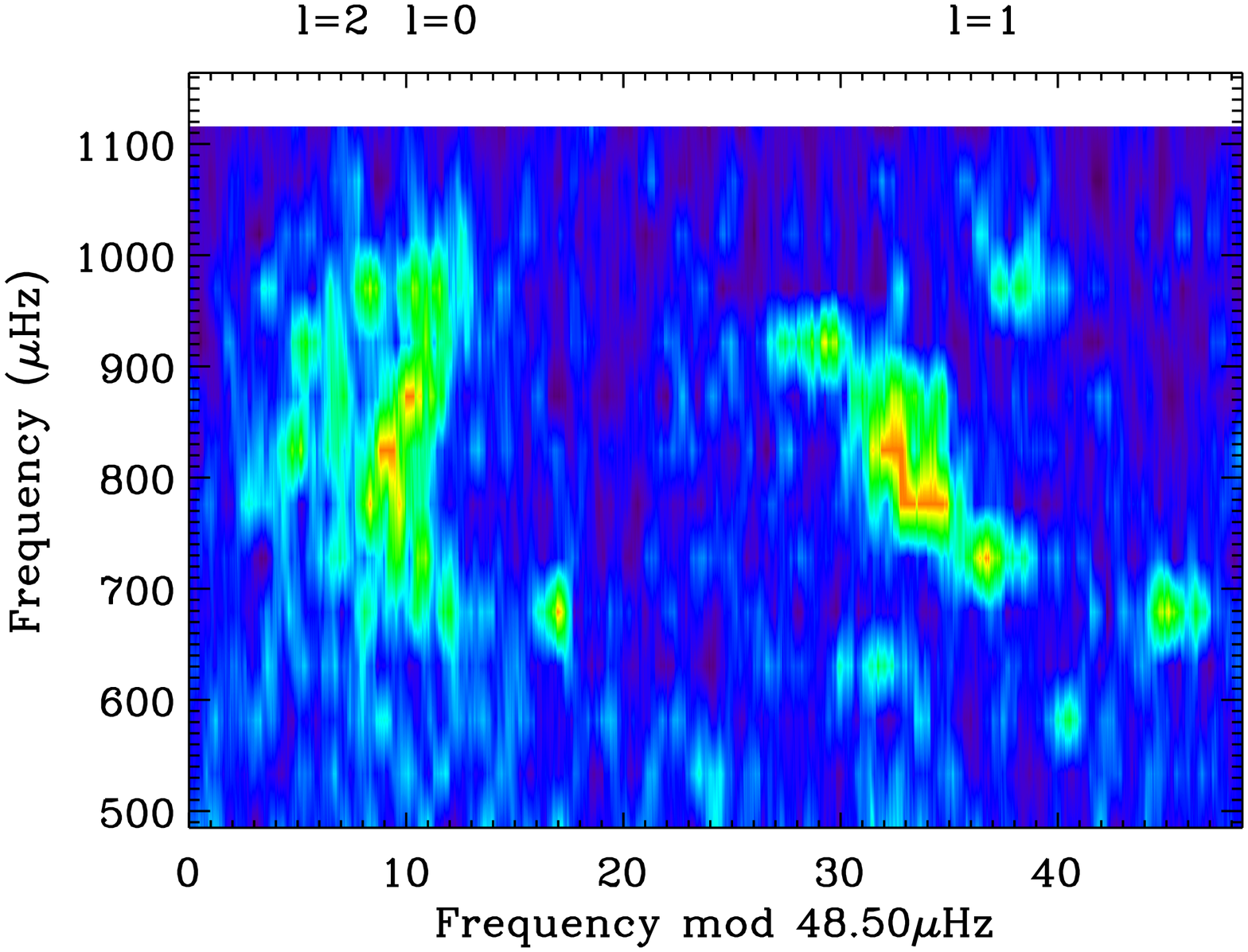}
\includegraphics[width=4cm, angle=90]{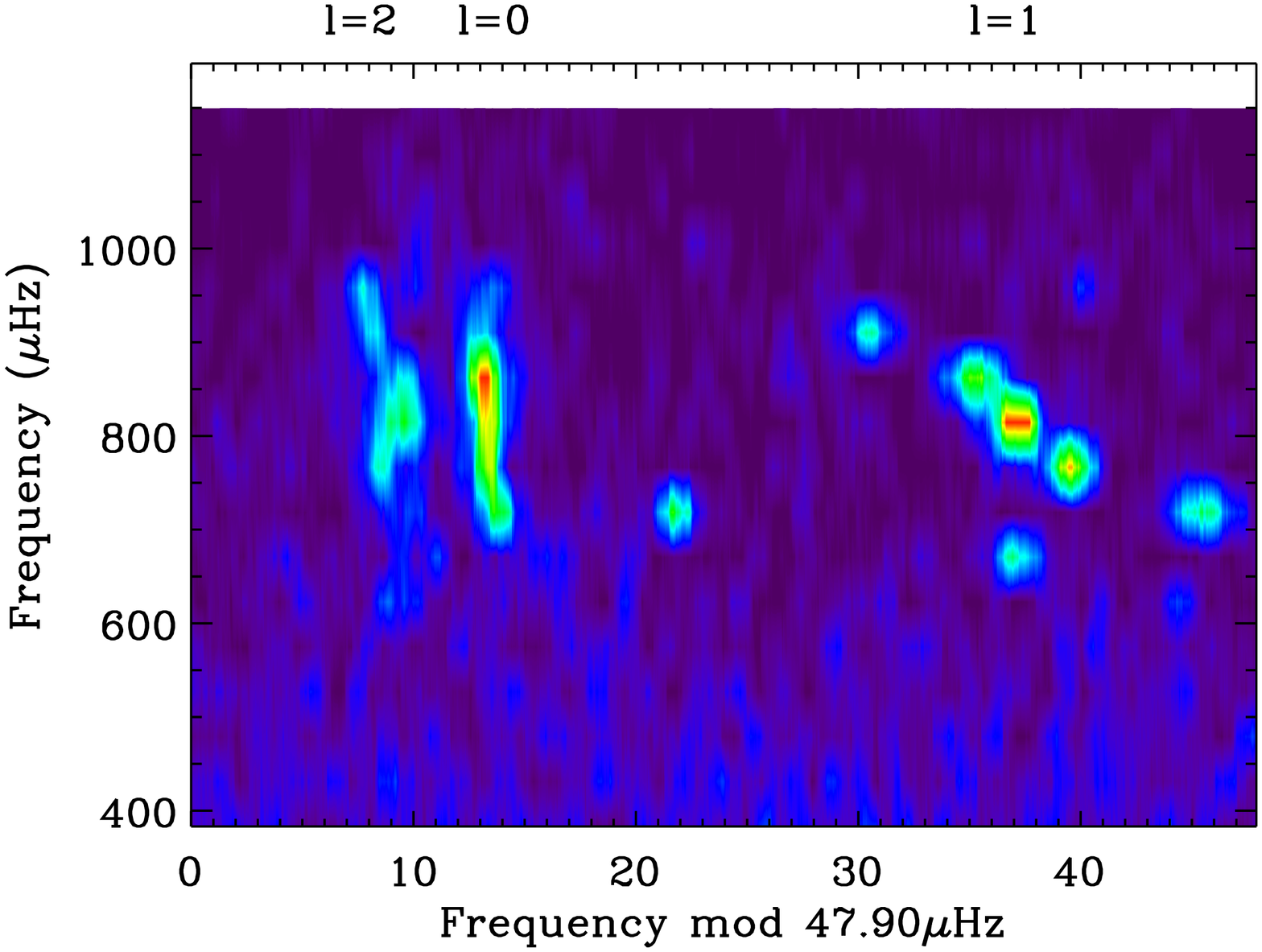}
\caption{Echelle diagram of KIC~10273246 (left panel) and KIC~11395018 (right panel) obtained with more than eight months of {\it Kepler} data. We can see the ridges $\ell$\,=\,0, 1, and 2 and the presence of the mixed modes (see the text for more details).}
\label{fig1}
\end{center}
\end{figure*}

Several groups inside KASC analyzed the datasets to retrieve the global parameters (mean large frequency separation $\langle \Delta \nu \rangle$, frequency of maximum power $\nu_{\rm max}$, small separation $\langle \delta_{02} \rangle$, mean linewidth $\langle \Gamma \rangle$) as well as the p-mode characteristics. Figure~\ref{fig1} represents the \'echelle diagrams for two of these stars. We can see the presence of mixed modes with the so-called avoided crossing where modes are ``bumped''. Mixed modes are very interesting as the evolutionary stage of a star is reflected in their characteristics so they bring strong constraints on the stellar interior and on the age of the star \citep{2010ApJ...723.1583M}. From the global seismic parameters  and knowing the effective temperature from the {\it Kepler} Input Catalogue \citep[KIC,][]{2011arXiv1102.0342B} it is possible to have a first determination of the mass and the radius of the star using the scaling relations from solar values as defined by \citet{1995A&A...293...87K}. In Table~1 we summarize the seismic and inferred global parameters of the stars. We used $T_{\rm eff}$\,=\,5660\,K, 6240\,K, 6380\,K, and 5880\,K for KIC~11395918, KIC~11234888, KIC~10273246, and KIC~10920273 respectively.

\begin{table*}
\begin{center}
\caption{Global seismic parameters and inferred stellar properties.}
\begin{tabular}{ccccc}
\tableline\tableline
Star &  $\langle \Delta \nu \rangle$ ($\mu$Hz)&  $\nu_{\rm max}$ ($\mu$Hz) &  Mass/$M_\odot$ & Radius/$R_\odot$\\
\tableline
KIC~11395018 & 47.8 ~$\pm$~1.0 & 830~$\pm$~48 & 1.25~$\pm$~0.24 & 2.15~$\pm$~0.21 \\
KIC~11234888 & 41.7~$\pm$~0.9 & 675~$\pm$~42  &  1.33~$\pm$~0.26 & 2.40~$\pm$~0.24 \\
KIC~10273246 & 48.2~$\pm$~0.5 & 839~$\pm$~51  &   1.49~$\pm$~0.28 & 2.27~$\pm$~0.23\\
KIC~10920273 & 57.3~$\pm$~0.8 & 1024~$\pm$~64    &1.20~$\pm$~0.24 & 1.88~$\pm$~0.18\\
\tableline
\end{tabular}
\end{center}
\end{table*}

\section{Conclusions}
For the first time, we have more than eight months of continuous asteroseismic data for four solar-like stars. Unfortunately, KIC~11234888 and KIC~10920273 have not been observed after the fourth Quarter of the {\it Kepler} operations. In the case of KIC~11395018, 22 months will be available with an interruption of around two months due to the failure in the outputs of the third CCD-module. Finally, KIC~10273246, has not been observed in Quarters 6 and 7 but it has been put in the KASC target list from Quarter 8. With the data acquired, we first obtained global seismic parameters such as: $\nu_{\rm max}$, $\langle \Delta \nu \rangle$, $\langle \delta_{02}\rangle$, $\langle \Gamma \rangle$ as well as other parameters like orbital period: $P_{\rm rot}$. Using the scaling relations from solar values, we could provide a first model-independent estimation of the mass and radius of these four stars. Then, it was also possible to analyze individual modes. We identified 16 to 30 modes for each star. 
Detailed modeling is now on going using all these seismic constraints coupled with atmospheric parameters. This will allow to check to a very precise level the physical process involved in the stellar interiors (Creevey et al. submitted; Brand\~ao et al. in prep.; Do\u{g}an et al. in prep.).

\acknowledgements 
Funding for this Discovery mission is provided by NASAs Science Mission Directorate. The authors wish to thank the entire Kepler team, without whom these results would not be possible. We also thank all funding councils and agencies that have supported the activities of KASC Working Group 1, and the International Space Science Institute (ISSI). NCAR is supported by the National Science Foundation. SH also acknowledges financial support from the Netherlands Organization for Scientific Research (NWO). JM\.Z acknowledges the Polish Ministry grant no N\,N203\,405139.\\

\noindent Affiliations: 
{$^1$High Altitude Observatory, NCAR, P.O. Box 3000, Boulder, CO 80307, USA.}
{$^2$Centro de Astrof\'isica, DFA-Faculdade de Ci\^encias, Universidade do Porto, Rua das Estrelas, 4150-762 Porto, Portugal.}
{$^3$Department of Physics and Astronomy, Aarhus University, DK-8000 Aarhus C, Denmark.}
{$^4$La\-bo\-ra\-toi\-re AIM, CEA/DSM -- CNRS - Universit\'e Paris Diderot -- IRFU/SAp, 91191 Gif-sur-Yvette Cedex, France.}
{$^5$Institut d'Astrophysique Spatiale, UMR8617, Universit\'e Paris XI, Batiment 121, 91405 Orsay Cedex, France.}
{$^6$Sydney Institute for Astronomy, School of Physics, University of Sydney, NSW 2006, Australia.}
{$^7$LESIA, UMR8109, Universit\'e Pierre et Marie Curie, Universit\'e Denis Diderot, Obs. de Paris, 92195 Meudon Cedex, France.}
{$^8$School of Physics and Astronomy, University of Birmingham, Edgbaston, Birmingham B15 2TT, UK.}
{$^9$Institut de Recherche en Astrophysique et Plan\'etologie, Universit\'e de Tou\-louse, CNRS, 14 avenue E. Belin, 31400 Toulouse, France.}
{$^{10}$Universit\'e de Toulouse, UPS-OMP, IRAP, 31400 Toulouse, France.}
{$^{11}$INAF Osservatorio Astrofisico di Catania, Via S. Sofia 78, 95123, Catania, Italy.}
{$^{12}$Astro\-no\-mi\-cal Institute ``Anton Pannekoek'', University of Amsterdam, PO Box 94249, 1090 GE Amsterdam, The Netherlands.}
{$^{13}$Universidad de La Laguna, Dpto de Astrof\'isica, 38206, Tenerife, Spain.}
{$^{14}$Instituto de Astrof\'\i sica de Canarias, 38205, La Laguna, Tenerife, Spain.}
{$^{15}$Universit\'e de Nice Sophia-Antipolis, CNRS, Observatoire de la C\^ote dÕAzur, BP 4229, 06304 Nice Cedex 4, France.}
{$^{16}$Australian Astronomical Observatory, PO Box 296, Epping NSW 1710, Australia.}
{$^{17}$Departamento de F\'isica e Astronomia, Faculdade de Ci\^encias da Universidade do Porto, Portugal.}
{$^{18}$Institute of Astronomy, University of Vienna, A-1180, Vienna, Austria.}
{$^{19}$Astronomical Institute, University of Wroc{\l}aw, ul. Kopernika 11, 51-622 Wroc{\l}aw, Poland.}
{$^{20}$Space Telescope Science Institute, Baltimore, MD 21218, USA.}
{$^{21}$Department of Physics and Astronomy, Iowa State University, Ames, IA 50011, USA.}
{$^{22}$SETI Institute/NASA Ames Research Center, Moffett Field, CA 94035, USA.}
{$^{23}$NASA Ames Research Center, Moffett Field, CA 94035, USA.}
\bibliographystyle{asp2010} 
\bibliography{./BIBLIO.bib}

\end{document}